\documentclass[aps,prc,times,graphicx,showpacs]{revtex4}

\usepackage{amsmath}
\usepackage{amsfonts}
\usepackage{epsfig}
\usepackage{afterpage}

\allowdisplaybreaks[4]

\begin{document}

\def\M{M^{0,(RFG)}_A}
\def\MA{(M^0_{A-1})^2}
\def\E{{\cal E}}
\def\N{N_{\rm spin}^{\rm isospin}}
\def\up{u^\prime}
\def\vp{v^\prime}
\def\upsq{u^{\prime\, 2}}
\def\vpsq{v^{\prime\, 2}}


\title{A Model for BCS-Type Correlations in Superscaling}

\author{M.B. Barbaro$^a$, R. Cenni$^b$, T.W. Donnelly$^c$ and A. Molinari$^a$}

\affiliation{$^a$Dipartimento di Fisica Teorica, Universit\`a di Torino and
  INFN, Sezione di Torino, Via P. Giuria 1, 10125 Torino, ITALY}
\affiliation{$^b$INFN, Sezione di Genova, Via Dodecaneso 33, 16146 Genova,
ITALY}
\affiliation{$^c$Center for Theoretical Physics, Laboratory for Nuclear
  Science and Department of Physics, Massachusetts Institute of Technology,
  Cambridge, MA 02139, USA}




\begin{abstract}

Using ideas from BCS descriptions of systems of fermions, a
covariant extension of the relativistic Fermi gas model is presented
as a way to incorporate correlation effects in nuclei. The model is
developed for the BCS nuclear ground state and for final states
consisting of a single plane-wave nucleon plus a BCS recoiling
daughter nucleus. The nuclear spectral function is obtained and from
it the superscaling function for use in treating high-energy
quasielastic electroweak processes. Interestingly, this model has
the capability to yield the asymmetric tail seen in the experimental
scaling function.

\end{abstract}




\maketitle


\section{Introduction}
\label{sec:intro}

The issue of superscaling in electroweak interactions with atomic
nuclei has been the focus in recent studies~\cite{Alberico:1988bv,
Day,CDM,rely,DS199,DS299,Maieron:2001it,Barbaro:2003ie,nu,
Caballero:2005sj,Amaro:2005dn,Amaro:2006pr,Amaro:2006tf,
Caballero:2007tz,Martinez:2008ve,Martinez:2008ev,Martini:2007jw}
because of its intrinsic interest and because it offers a way to
obtain reliable charge-changing and neutral-current neutrino-nucleus
cross sections. The latter are usually essential when addressing
fundamental neutrino properties, specifically neutrino masses and
the neutrino oscillations that result from those masses. A summary
of the basic concepts involved in superscaling analyses is provided
below in Section~\ref{sec:long}.

Owing to the complexity of nuclear dynamics it is not obvious that
the nuclear response to an electroweak field superscales; indeed for
this to occur the single-nucleon cross section should factorize,
which is almost never completely true.
Certainly in extreme
on-shell independent-particle models such as the Relativistic Fermi
Gas (RFG) this is valid, although in general factorization is seen
to be violated to some degree, since the nucleons in the nucleus are
not on-shell and the independent-particle model is only an
approximation. For instance, in other studies done within the
context of the shell model (for instance, \cite{Caballero:2005sj,
Amaro:2005dn}) one sees breaking of factorization from off-shell
effects, typically at the few percent level for the kinematics of
interest when applying the ideas of superscaling, namely, momentum
transfers of order 1 GeV/c and energy transfers near the
quasielastic peak position. Clearly, if collective effects (for
instance, giants resonances) are important, the independent-particle
description breaks down and thus superscaling should not be expected
to hold in the low momentum and energy transfer domain where such
effects occur. Finally, when meson-exchange currents play a
significant role the superscaling violations that may or may not
occur require detailed
study~\cite{Amaro:2002mj,Amaro:2003yd,DePace:2003,DePace:2004}.

In the present work we shall not address these general issues, but
instead shall focus on an extension of the RFG approach, namely, we
shall include only on-shell nucleons in an independent-particle
model where factorization is exact and put our efforts into going
beyond the degenerate description provided by the extreme RFG model.
To do this we resort to a model inspired by the BCS theory of
condensed matter physics (see, {\it e.g.,} \cite{FW}) in which
electron Cooper pairs are correlated via an attractive interaction
mediated by the lattice. In nuclei correlations among pairs of
nucleons also occur~\cite{Rowe}, in this case mediated by NN
interactions including both the long-range attractive residual
interaction (the pairing force) and also the strong repulsive force
that occurs at short distances. However, and remarkably, the BCS
wave function can be shown to hold valid (in the thermodynamic
limit) in both instances in the sense of being an eigenstate of the
well-known pairing Hamiltonian independently of the sign of the
interaction (of course in the attractive case the BCS state provides
a minimum for the system's energy, in the repulsive one it yields
instead a maximum).

Hence we shall deal with the problem of assessing the impact of
correlations on the superscaling function employing the BCS
formalism with appropriate modifications, such as retention of
covariance, to adapt it to the high-energy physics of atomic nuclei.
This is addressed in Section~\ref{sec:SF} after having briefly
outlined in Section~\ref{sec:long} the basics of the superscaling
approach for the longitudinal nuclear response $R_L$, namely the
part of the total inclusive electroweak response that is believed to
superscale the best.

The numerical results obtained with our approach are presented in
Section~\ref{sec:results}. To summarize briefly our findings, we
shall see that, likely because in the BCS spirit we limit ourselves
to an independent quasi-particle description of nuclear matter,
scaling of the first kind (independence of momentum transfers $q$)
appears to occur not only at the quasielastic peak (QEP), but also
at both lower and higher energy transfers $\omega$. We find that the
onset of first-kind scaling already occurs at momentum transfers of
order 500 MeV/c, whereas away from the QEP the onset only occurs at
quite large momentum transfers (of the order of 2 GeV/c).
Furthermore, interestingly the shape of the superscaling function
turns out to be {\em non-symmetric} around the QEP, being larger to
the right and smaller to the left of it, namely, in agreement with
experiment and thus lending support to our approach. However, in our
model when in the so-called scaling region (below the QEP)
first-kind scaling is reached as a function of $q$ from below,
which is not what is
experimentally found. Finally, scaling of the second kind
(independence of nuclear species) is also explored in
Section~\ref{sec:results} and shown to be relatively well satisfied,
given that an appropriate momentum scale is chosen for each nuclear
species.

In the Conclusions (Section~\ref{sec:concl}) we briefly discuss the
significance and limitations of our model.


\section{The longitudinal response function}
\label{sec:long}

As anticipated in the Introduction we shall concentrate on the
longitudinal EM channel; all other electroweak responses can be
developed using similar arguments to those presented in the
following. We start by briefly summarizing the basics of the
superscaling approach, which has been discussed in depth in previous
work~\cite{Alberico:1988bv,
Day,CDM,rely,DS199,DS299,Maieron:2001it,Barbaro:2003ie,nu,
Caballero:2005sj,Amaro:2005dn,Amaro:2006pr,Amaro:2006tf,
Caballero:2007tz,Martinez:2008ve,Martinez:2008ev}.
Here we employ the same notation and formalism as in the paper by
Cenni-Donnelly-Molinari~\cite{CDM} (to be referred to as CDM).

Within the framework of the Plane-Wave Impulse Approximation
(PWIA) the longitudinal response function of a nucleus to an
external electroweak field bringing three-momentum $\bf q$ and
energy $\omega$ into the system reads
\begin{equation}
\label{eq:RL} R_L(q,\omega)=\int\!\!\!\!\int_\Sigma dp\,d{\cal
E}\,p^2\frac{E_N}{pq}2\pi S(p,{\cal
E})\frac{m_N}{E_p}\frac{m_N}{E_N} {\cal R}_L(q,\omega,p,{\cal
E})\, ,
\end{equation}
where $E_p=\sqrt{{\bf p}^2+m_N^2}$ and $E_N=\sqrt{({\bf p}+{\bf
q})^2+m_N^2}$ are the energies of the initial (struck) and final
(ejected) nucleons, respectively, with $m_N$ the nucleon mass.
Indeed in PWIA it is assumed that only one vector boson is
exchanged between the probe and the nucleus and that this one is
absorbed by a single nucleon. The probability of finding one
nucleon in the system is provided by the spectral function of the
system, $S(p,{\cal E})$, which depends on the missing momentum $p$
and an energy ${\cal E}$, the latter being essentially the missing
energy minus the separation energy and given explicitly by the
excitation energy of the residual nucleus in the reference frame
where it moves with momentum $-{\bf p}$. The factor ${\cal R}_L$
in Eq.~\eqref{eq:RL} is the longitudinal single-nucleon response
which is in general half-off-shell and hence a function not only
of $q$ and $\omega$, but also of the energy and momentum of the
off-shell struck nucleon, or equivalently of $p$ and ${\cal E}$.
In the models being considered in the present study the struck
nucleon is in fact on-shell and so ${\cal R}_L$ becomes simply the
longitudinal response of a moving free nucleon:
\begin{equation}
\label{eq:RLa} {\cal R}_L=\frac{\kappa^2}{\tau}\left\{G_E^2(\tau)+
W_2(\tau)
\chi^2\right\}\, ,
\end{equation}
where
\begin{equation}
\label{eq:RLb} W_2(\tau)=\frac{1}{1+\tau}\left[G_E^2(\tau)+\tau
G_M^2(\tau) \right]\, ,
\end{equation}
expressed in terms of the electric $G_E$ and magnetic $G_M$ Sachs
form factors of the nucleon and $\chi=(p/m_N)\sin\theta$, where
$\theta$ is the angle between $\bf p$ and $\bf q$. As in CDM in
Eqs.~\eqref{eq:RLa} and \eqref{eq:RLb} the dimensionless variables
\begin{equation}
\kappa=\frac{q}{2m_N}\,,\,\,\,
\lambda=\frac{\omega}{2m_N}\,\,\mbox{and}\,\,
\tau=\kappa^2-\lambda^2
\end{equation}
are employed.

Equation~\eqref{eq:RL} connects the semi-inclusive $(e,e'N)$
reaction with the inclusive $(e,e')$ process assuming that the
outgoing nucleon no longer interacts with the residual $(A-1)$
nucleus (absence of final-state interactions). That equation
expresses the assumption that the inclusive cross section is to be
obtained by integrating the semi-inclusive cross section, summing
over struck protons and neutrons. The boundaries of the integration
domain $\Sigma$ in the $({\cal E},p)$ plane are found through the
energy conservation relation (see CDM) and read
\begin{equation}
\label{eq:boundaries} {\cal E}^{\pm}(p) = \omega-E_s+m_N-\sqrt{(p\pm
q)^2+m_N^2}-T_0^{(A-1)}~,
\end{equation}
where ${\cal E}^{+} \leq {\cal E} \leq {\cal E}^{-}$  and ${\cal
E}^{-}$ is bounded according to
\begin{equation}
0 \leq {\cal E}^{-} \leq \omega -E_s -\left[ \sqrt{q^2
+(M_{A-1}+m_N)^2} - (M_{A-1}+m_N) \right]\, .
\end{equation}
In the above
\begin{equation}
\label{eq:sep}
E_s=M_{A-1}+m_N-M_A
\end{equation}
is the separation energy of a nucleon from the ground state of the
nucleus $A$ (of mass $M_A$) leaving the residual $(A-1)$ nucleus
(of mass $M_{A-1}$) in its ground state as well and
\begin{equation}
T_0^{(A-1)} = \sqrt{p^2+M_{A-1}^2}-M_{A-1}
\end{equation}
is the very small recoiling nucleus kinetic energy, which is often
neglected in Eq.~\eqref{eq:boundaries}. Careful investigations of
Eq.~\eqref{eq:RL} have shown that the single-nucleon response
function is quite gently varying over the integration domain
$\Sigma$ in the $({\cal E},p)$ plane, particularly if one limits
the focus only to regions where the spectral function $S(p,{\cal
E})$ plays a significant role (see CDM); accordingly it can be
extracted from the integral in Eq.~\eqref{eq:RL} and divided out,
thereby incurring only typically few percent errors. This is the
point at which the factorization discussed in the Introduction is
invoked.

Finally, in this study we shall confine ourselves to dealing with
infinite, homogeneous systems, the simplest among them being the RFG
model in which the dynamics are controlled by just one parameter,
the Fermi momentum $k_F$. To attain superscaling it then turns out
to be convenient to multiply Eq.~\eqref{eq:RL} by a further factor
that, in the RFG case, is proportional to the momentum transfer and
to $k_F$. Indeed, as we shall see, this leads to a function which
loses any dependence on both $k_F$ and $q$, namely, one has
superscaling for the RFG in the non-Pauli-blocked regime. It remains
to be seen what happens in the BCS model, {\it i.e.,} in the
presence of correlations.


\section{The spectral function}
\label{sec:SF}

As discussed in the previous section the key tool for exploring
superscaling is the nuclear spectral function $S(p,{\cal E})$,
defined as follows
\begin{equation}
  S(p,{\cal E})=<\Psi_0|a^\dagger_{{\bf p}\uparrow}\delta({\cal E}
  -({\hat H}-E_0^{(A-1)}))a_{{\bf p}\uparrow}|\Psi_0> \frac{V_A}{A(2\pi)^3}
\end{equation}
in terms of matrix elements of contributions involving the
standard fermion creation and annihilation operators. In the above
$|\Psi_0>$ and $\hat H$ are the ground state and hamiltonian of
the system, respectively, and $V_A$ is the (large) volume
enclosing the system.


\subsection{The RFG model}

The RFG spectral function is easily computed when one recalls (see
CDM) that it has a negative separation energy
\begin{equation}
\label{eq:EsRFG}
(E_s)_{RFG}=-T_F~,
\end{equation}
$T_F=\sqrt{k_F^2+m_N^2}-m_N$ being the Fermi kinetic energy. One
obtains
\begin{equation}
\label{eq:SRFG}
S^{RFG}(p,{\cal E}) = \sum_{spin,isospin} \theta(k_F-p) \delta({\cal
  E}-T_F+T_p) \frac{V_A}{A(2\pi)^3}~,
\end{equation}
where the nucleon kinetic energy $T_p=\sqrt{p^2+m_N^2}-m_N$ has
been introduced, and observes that it is nonzero in the $({\cal
E},p)$ plane only on the curve
\begin{equation}
{\cal E}=\sqrt{k_F^2+m_N^2}-\sqrt{p^2+m_N^2}
\label{eq:dispRFG}
\end{equation}
and is normalized to unity, as is easy to check by integrating
Eq.~\eqref{eq:SRFG} over ${\bf p}$ and the non-negative variable
${\cal E}$. An integration only over ${\cal E}$ yields the system's
momentum distribution normalized to the Fermi sphere, namely
\begin{equation}
\int_0^\infty d{\cal E} S^{RFG}(p,{\cal E}) =
\frac{\theta(k_F-p)}{\frac{4}{3}\pi k_F^3}~.
\end{equation}

Inserting Eq.~\eqref{eq:SRFG} into Eq.~\eqref{eq:RL}, after having
pulled out and divided by the single-nucleon response ${\cal
R}_L$, one obtains
\begin{equation}
\label{eq:f} F(q,\omega) = \frac{2\pi m_N^2}{q}
\int\!\!\!\!\int_{\Sigma_{RFG}} dp\, d{\cal E} \frac{p}{E_p}
S^{RFG}(p,{\cal E})~,
\end{equation}
where the domain $\Sigma_{RFG}$ is defined by the curves in
Eq.~\eqref{eq:boundaries} using Eq.~\eqref{eq:EsRFG}. Performing
the integration, which can be done analytically, and dividing by
the factor
\begin{equation}
\label{eq:K} \Lambda =
\frac{1}{k_F}\left(\frac{m_N}{q}\right)\left(\frac{2m_N
T_F}{k_F^2}\right)
\end{equation}
(note that the quantity $2m_N T_F/k_F^2$ is very close to unity),
one ends up with the RFG superscaling function~\cite{Alberico:1988bv}
\begin{equation}
\label{eq:fRFG} f(\psi) = \frac{3}{4} \left(1-\psi^2\right) \theta
\left(1-\psi^2\right)~,
\end{equation}
where
\begin{equation}
\label{eq:psi} \psi = \frac{1}{\sqrt{\xi_F}}
\frac{\lambda-\tau}{\sqrt{(1+\lambda)
\tau+\kappa\sqrt{\tau(1+\tau)}}}
\end{equation}
with $\xi_F=T_F/m_N$. Here $\psi$ is the scaling variable
specifically obtained within the context of the RFG, although it has
also been widely employed in more general analyses.


\subsection{The BCS-inspired model}

In the spirit of BCS theory we assume that the ground state is now
given by
\begin{equation}
\label{eq:BCS}
  |BCS>=\prod_k(u_k+v_k a^\dagger_{k\uparrow}a^\dagger_{-k\downarrow})|0>~,
\end{equation}
$|0>$ being the true vacuum, with
\begin{equation}
\label{eq:normBCS}
|u_k|^2+|v_k|^2=1~,
\end{equation}
to insure the proper normalization of the state, namely
\begin{equation}
<BCS|BCS>=\prod_k(|u_k|^2+|v_k|^2)=1~.
\end{equation}
With the assumption in Eq.~\eqref{eq:BCS} we have a covariant
approximation to the nuclear ground-state wavefunction. We have
required that the added pairs always occur with back-to-back momenta
(hence the net linear momentum of the system in its rest frame is
still zero) and with opposite helicities (hence the net spin of the
ground state is zero; we consider only even-even nuclei for
simplicity). The creation operators add particles with relativistic
on-shell spinors $u(k,\lambda)$.

The state $|BCS>$ does not correspond to a fixed number of
particles; in fact it is associated with the spontaneous breaking of
the $U(1)$ symmetry in the system. Therefore, it is not an
eigenstate of the operator
\begin{equation}
\label{eq:opnum} \hat n(k)=\sum_s a^\dagger_{k s}a_{k s}\, ,
\end{equation}
although its expectation value can be computed, yielding
\begin{equation}
  n(k)=<BCS|\hat n(k)|BCS>=|v_k|^2~.
\end{equation}
We can thus require the particle number $A$ to be conserved on the
average, which implies the condition
\begin{equation}
  \sum_k |v_k|^2=A~.
\label{eq:npar}
\end{equation}

Concerning the energy, we view our system as being constructed in
terms of independent quasi-particles, which indeed is the case for
the state in Eq.~\eqref{eq:BCS} in leading order, writing
accordingly
\begin{equation}
\label{eq:EBCS}
  E_{BCS} = <BCS|\hat H|BCS>=<BCS|\sum_{ks}E_k
  a^\dagger_{ks}a_{ks}|BCS>=\sum_{ks} E_k |v_k|^2~.
\end{equation}
For sake of simplicity in the following we make the assumption
\begin{equation}
  E_k=\sqrt{m_N^2+k^2}~,
\end{equation}
which reflects the fact that our present goal is to obtain a
reasonable variational wave function (the BCS one), not for the
purpose of getting the best energy, but rather to provide a way to
evaluate the role of correlations on superscaling.

We now face the question of how to compute the wave functions of the
daughter nucleus. For this we assume
\begin{equation}
\label{eq:daugh}
  |D(p)>=\frac{1}{\sqrt{{\cal N}_p}}\,a_{p\uparrow}|BCS^\prime>\,
  ,
\end{equation}
where ${\cal N}_p$ is a normalization factor, and
\begin{equation}
  |BCS^\prime>=\prod_k[\up_k(p)+\vp_k(p)
   a^\dagger_{k\uparrow}a^\dagger_{ -k\downarrow}]|0>~,
\end{equation}
where the coefficients $\up_k(p)$ and $\vp_k(p)$ are in general
different from those in Eq.~\eqref{eq:BCS}, but still obey
Eq.~\eqref{eq:normBCS}. This point is of crucial relevance for our
model, as we shall see below. Note that the daughter nucleus states
are labeled by the momentum, {\bf p}. For the normalization factor
in Eq.~\eqref{eq:daugh} we get
\begin{equation}
  {\cal N}_p=|\vp_p(p)|^2
\end{equation}
and the expectation value of the operator in Eq.~\eqref{eq:opnum}
(the system's momentum distribution) turns out to be 
(see Appendix~\ref{sec:appB})
\begin{equation}
  n_{D(p)}(k)=<D(p)|\sum_sa^\dagger_{ks}a_{ks}|D(p)>
  =|\vp_k(p)|^2 (1-\delta_{kp})~.
\end{equation}
Clearly the condition on the particle number will now read
\begin{equation}
\label{eq:ndau}
\sum_{k\not=p}|\vp_k(p)|^2=\sum_{k}|\vp_k(p)|^2-|\vp_p(p)|^2
=A-1~,
\end{equation}
valid for all of the daughter states, namely for all $p$. Taking
into account Eq.~\eqref{eq:npar}, this yields the relation
\begin{equation}
\label{eq:vvp}
\sum_{k}|v_k|^2-\sum_{k}|\vp_k(p)|^2
=1-|\vp_p(p)|^2~,
\end{equation}
to be exploited below.

The energy of the daughter nucleus in our model is
\begin{equation}
E_{D(p)} = <D(p)|\hat H|D(p)> = \sum_{k\not=p} E_k |\vp_k(p)|^2~,
\end{equation}
which, using Eqs.~\eqref{eq:EBCS} and \eqref{eq:vvp}, can be recast
as follows:
\begin{equation}
  \label{eq:Edau}
E_{D(p)} = (E_{BCS}-m_N)-T_p |v_p|^2-\sum_{k\not=p} T_k
[|v_k|^2-|\vp_k(p)|^2]~.
\end{equation}
We can then proceed to compute the daughter nucleus spectral
function
\begin{equation}
\label{eq:SFBCS}
  S^{BCS}(p,{\cal E})=\left|<D(p)|a_{p\uparrow}|BCS>\right|^2
\delta\left[{\cal E}-\left(E_{D(p)}-E_0^{(A-1)}\right)\right]
\frac{V_A}{A(2\pi)^3}~,
\end{equation}
where $E_0^{(A-1)}$ is the energy $E_{D(p)}$ of the daughter
nucleus evaluated at that value of $p$ where it reaches its
minimum, to be referred to as $k_F$ in the BCS model. Hence we
have
\begin{eqnarray}
{\cal E}(p) &=& E_{D(p)}-E_{D(k_F)}
\nonumber\\
&=&
T_F |v_{k_F}|^2 - T_p |v_p|^2
+\sum_{k\not=k_F} T_k [|v_k|^2-|\vp_k(k_F)|^2]
-\sum_{k\not=p} T_k [|v_k|^2-|\vp_k(p)|^2]~.
\label{eq:cale}
\end{eqnarray}
Later we shall show that the last two terms on the right-hand side
of the above provide only small corrections and so can be dropped.
We thus see that our model provides a nice variation of the RFG
expression in Eq.~\eqref{eq:dispRFG}; see later where results are
displayed. The matrix element in
Eq.~\eqref{eq:SFBCS} can be straightforwardly computed (see
Appendix~\ref{sec:appC}) and one obtains
\begin{eqnarray}
{\cal M}_p(p) &=& \left|<D(p)|a_{p\uparrow}|BCS>\right|^2
\nonumber\\
&=& v_p \sqrt{\frac{\vp_p(p)}{\vp_p(p)^*}}
\prod_{k\not=p}[\up_k(p)^* u_k+\vp_k(p)^* v_k]~.
\end{eqnarray}
Thus we end up with the expression
\begin{eqnarray}
  S^{BCS}(p,{\cal E})&=&|{\cal M}_p(p)|^2 \delta[{\cal E}-{\cal
      E}(p)]\frac{V_A}{A(2\pi)^3}
\nonumber\\
&=& |v_p|^2 \delta[{\cal E}-{\cal E}(p)]
\left|\prod_{k\not=p}[\up_k(p)^* u_k+\vp_k(p)^* v_k]\right|^2
\frac{V_A}{A(2\pi)^3}
\label{eq:SFBCS1}
\end{eqnarray}
for the daughter nucleus spectral function.

Finally, in order to calculate the longitudinal response function in
Eq.~\eqref{eq:RL} what remains to be specified is the integration
region $\Sigma$, which in turn requires knowledge of the separation
energy in Eq.~\eqref{eq:sep}. In the present model the latter turns
out to be
\begin{equation}
\label{eq:EsBCS} (E_s)_{BCS} = -T_F |v_{k_F}|^2 -
\sum_{k\not=k_F} T_k \left[|v_k|^2-|\vp_k(k_F)|^2\right]~.
\end{equation}

The question we are now faced with is how to fix the coefficients
$u_k$, $v_k$, $\up_k(p)$ and $\vp_k(p)$ entering in the wave
functions of the initial ground state and final daughter nucleus
state. As a first step in this direction let us recall the
normalization constraints that these coefficients have to fulfill
and deduce expressions for them in the thermodynamic limit
$A\to\infty$, $V_A\to\infty$, $A/V_A=\rho_A$. For the initial
nucleus one has
\begin{equation}
\lim \frac{1}{V_A} \sum_k |v_k|^2 =
\int\frac{d^3k}{(2\pi)^3} |v(k)|^2 = \lim \frac{A}{V_A} =\rho_A~,
\label{eq:norma}
\end{equation}
while for the daughter nucleus one obtains
\begin{eqnarray}
\lim \frac{1}{V_{A-1}}\left[\sum_k |\vp_k(p)|^2
-|\vp_p(p)|^2 \right]
&=& \int\frac{d^3k}{(2\pi)^3} |\vp(k;p)|^2 -
\left(\rho_A\frac{V_A}{V_{A-1}}-\rho_{A-1}\right)|\vp(p;p)|^2
\nonumber\\
&=& \lim \frac{A-1}{V_{A-1}}= \rho_{A-1}~.
\label{eq:normb}
\end{eqnarray}
Note that if Eq.~\eqref{eq:norma} is valid then the spectral
function in Eq.~\eqref{eq:SFBCS1} is correctly normalized, namely
$\int d^3p d{\cal E} S^{BCS}(p,{\cal E}) = 1$, provided that
$\prod_{k\not=p}[\up_k(p)^* u_k+\vp_k(p)^* v_k]=1$.

The next constraint for the daughter nucleus wave function's
coefficients stems from the stability condition, which requires
that the energy of the latter fulfill the relation
\begin{equation}
\label{eq:stab}
\left.\frac{d E_{D(p)}}{dp}\right|_{p=k_F}=0~,
\end{equation}
which, as already mentioned, defines the Fermi momentum $k_F$ for
the $(A-1)$ system. Neglecting the last term on the right-hand
side of Eq.~\eqref{eq:Edau} this becomes
\begin{equation}
\label{eq:stabil} \left.  \frac{d}{dp} \left[T(p) |v(p)|^2\right]
\right|_{p=k_F} =0\, ,
\end{equation}
which can be recast as follows
\begin{equation}
|v(k_F)|^2 \left.  \frac{d T(p) }{d p} \right|_{p=k_F}
+ T(k_F) v(k_F)^* \left.  \frac{d v(p) }{d p} \right|_{p=k_F}
+ T(k_F) v(k_F) \left.  \frac{d v(p)^* }{d p} \right|_{p=k_F}
=0~.
\end{equation}
Assuming that $v$ is real, this becomes
\begin{equation}
\label{eq:stab1} v(k_F) \frac{k_F}{\sqrt{k_F^2+m_N^2}} + 2
\left(\sqrt{k_F^2+m_N^2}-m_N\right) \left.  \frac{d v(p) }{d p}
\right|_{p=k_F} =0~.
\end{equation}

In accord with the BCS model and with the physics we intend to
explore in the next section, we now choose the following
three-parameter expression
\begin{equation}
  v^2(k)=\frac{c}{e^{\beta(k-\tilde k)}+1}
\label{eq:v2}
\end{equation}
for the $v$ coefficients. Note that when $\beta \to \infty$ one must
recover the RFG. Substituting Eq.~\eqref{eq:v2} into
Eq.~\eqref{eq:stab1} we obtain
\begin{equation}
  k_F\left[e^{\beta(k_F-\tilde k)}+1\right]=\beta\sqrt{k_F^2+m_N^2}
\left(\sqrt{k_F^2+m_N^2}-m_N\right) e^{\beta(k_F-\tilde k)}~,
\end{equation}
which can be solved for $\tilde k$ in terms of $k_F$ and $\beta$,
yielding
\begin{equation}
\label{eq:ktilde}
  \tilde k=k_F+\frac{1}{\beta}\log\left[\frac{\beta}{k_F}
\sqrt{k_F^2+m_N^2} \left(\sqrt{k_F^2+m_N^2}-m_N\right)-1\right]~.
\end{equation}
The inverse of this equation then gives $k_F$ in terms of $\tilde k$
and $\beta$.

We use next the normalization condition in Eq.~(\ref{eq:norma}) to
fix the parameter $c$ in Eq.~\eqref{eq:v2}. We obtain
\begin{equation}
\label{eq:c1}
c(\beta,\tilde k) = -\frac{\pi^2\beta^3\rho_A}{
  Li_3\left(-e^{\beta\tilde k}\right)}\, ,
\end{equation}
where
\begin{equation}
\label{eq:I}
Li_3(z) = \frac{1}{2} \int_0^\infty \frac{k^2}{e^k/z-1}\, dk
\end{equation}
is the trilogarithmic function.

For the daughter nucleus we employ the same form of
parametrization as in Eq.~\eqref{eq:v2}, namely
\begin{equation}
\label{eq:v2prime}
  \vpsq(k;p)=\frac{c_D}{e^{\beta_D\left[k-\tilde k_D(p)\right]}+1}
\end{equation}
and require the parameters to fulfill the normalization condition
for the $A-1$ system, namely, Eq.~\eqref{eq:normb}. Assuming
$\rho_{A-1}=\rho_A\equiv\rho$ and $V_A/V_{A-1}=A/(A-1)$ this leads
to the equation
\begin{equation}
\frac{c}{\pi^2\beta^3}Li_3\left(-e^{\beta\tilde k}\right) +
\frac{c_D}{\pi^2\beta_D^3}Li_3\left(-e^{\beta_D\tilde k_D(p)}\right) =
\frac{\rho}{A-1} \times \frac{c_D}{e^{\beta_D[p-\tilde k_D(p)]}+1}~.
\label{eq:eqkprimetilde}
\end{equation}
In the thermodynamic limit the right-hand side vanishes and
Eq.~\eqref{eq:eqkprimetilde} is simply solved by $c_D=c$,
$\beta_D=\beta$, $\tilde k_D(p)=\tilde k$; hence
$\vpsq(k;p)=v^2(k)$. As anticipated, this implies that the last two
terms in Eq.~\eqref{eq:cale} vanish and that the product in
Eq.~\eqref{eq:SFBCS1} is unity, so that the spectral function of the
daughter nucleus is correctly normalized to unity. It must be
emphasized that the coefficients $v$ and $v^\prime$ become identical
in the thermodynamic limit, but are different for finite $A$. Hence
it is crucial to compute the dispersion relation ${\cal E}(p)$ when
$A$ is finite and {\em then} take the thermodynamic limit. This is
reminiscent of Koopmans' theorem in Hartree-Fock theory~\cite{Koopmans}.

As far as the parameter $\beta$ is concerned, it clearly controls
both the modifications of the momentum distribution near the Fermi
surface (promotion of pairs due to residual NN interactions, both
long- and short-range) and also the tail of the momentum
distribution due to short-range NN correlations.
Indeed, for $\beta$ very large one
recovers the familiar $\theta$-distribution, while for smaller and
smaller $\beta$ more and more particles are pulled out of the Fermi
sea and produce a significant tail for the momentum distribution at
large momenta. The impact of the physics expressed by the parameter
$\beta$ on the superscaling function is explored in the next
section.


\section{Numerical results}
\label{sec:results}

The model is specified by three parameters, $\tilde k$, $\beta$ and
$c$, the last being fixed by the normalization condition in
Eq.~(\ref{eq:c1}). The Fermi momentum in this model is determined by
the daughter nucleus stability condition in Eq.~(\ref{eq:stabil}).
Alternatively, as it is somewhat easier to achieve, one can fix the
Fermi momentum $k_F$ and use Eq.~(\ref{eq:ktilde}) to determine
$\tilde k$.
In presenting the results obtained using our model it is then
convenient to start by displaying the behaviour of the parameters
$\tilde k$ and $c$, which are fixed by the physical conditions of
normalization and stability, versus $\beta$ for given $k_F$. When
$\tilde k$, $c$ and $\beta$ are known so are the wave functions of
the initial and final nuclei.

\begin{figure}
\label{fig:fig1}
\includegraphics[scale=0.8]{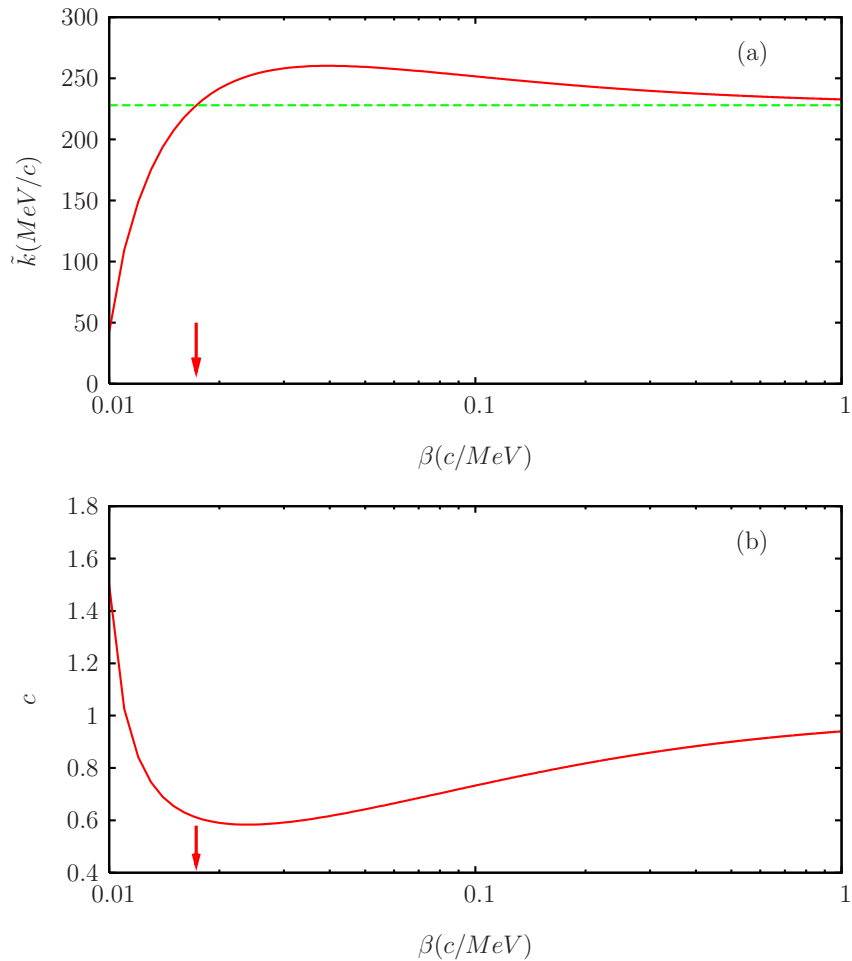}
\caption{(Color online) 
Panel (a):
The parameter $\tilde k$, given in Eq.~(\ref{eq:ktilde}),
shown as a function of $\beta$. The straight line indicates the Fermi 
momentum $k_F=228$ MeV/c.
Panel (b):
The parameter $c$, given in Eq.~(\ref{eq:c1}), plotted versus 
$\beta$ for $\rho_A=k_F^3/(6\pi^2)$.
In both panels the arrow indicates the critical value
$\beta_{\rm crit}= 0.017$ c/MeV.}
\end{figure}
In Fig.~1 the parameters $\tilde k$ and $c$ are plotted
versus $\beta$. They stay constant (in fact the almost constant
value of $\tilde k$ is quite close to the value $k_F=$ 228 MeV/c
we use as an input) until a critical value $\beta_{\rm
  crit}=0.017$ c/MeV is reached where $\tilde k$ ($c$) displays a dramatic
decrease (increase). This value corresponds to the change of sign
of the logarithmic term in Eq.~\eqref{eq:ktilde}, namely
\begin{equation}
\beta_{\rm crit}=\frac{2 k_F}{T_F (T_F+m_N)}~. \label{eq:betacrit}
\end{equation}
For values of $\beta$ lower than $\beta_{\rm crit}$ the nuclear
momentum distribution becomes very much extended beyond the Fermi
sphere associated with the input value of $k_F$; we shall return
below to discuss this delicate issue. Thus our results appear to
point to the existence of a narrow domain of $\beta$ around
$\beta_{\rm crit}$, below which the system becomes strongly
disrupted by correlations. This has a strong impact on the structure
of the superscaling function, as we shall see later. As stated
above, for the results shown in the figures we have taken
$\rho=k_F^3/(3\pi^2)$ and $k_F=228$ MeV/c (a value appropriate for
$^{12}$C), although our results appear to be quite insensitive to
$k_F$ as long as it is kept within the range appropriate for the
physics of atomic nuclei. We display the momentum distribution of
the initial nucleus in Fig.~2 for a few values of $\beta$ larger
(panel $a$) or smaller (panel $b$) than $\beta_{\rm crit}$. The
progressive development of a tail in the momentum distribution is
clearly seen in the figure.

\begin{figure}
\label{fig:fig3}
\includegraphics[scale=0.8]{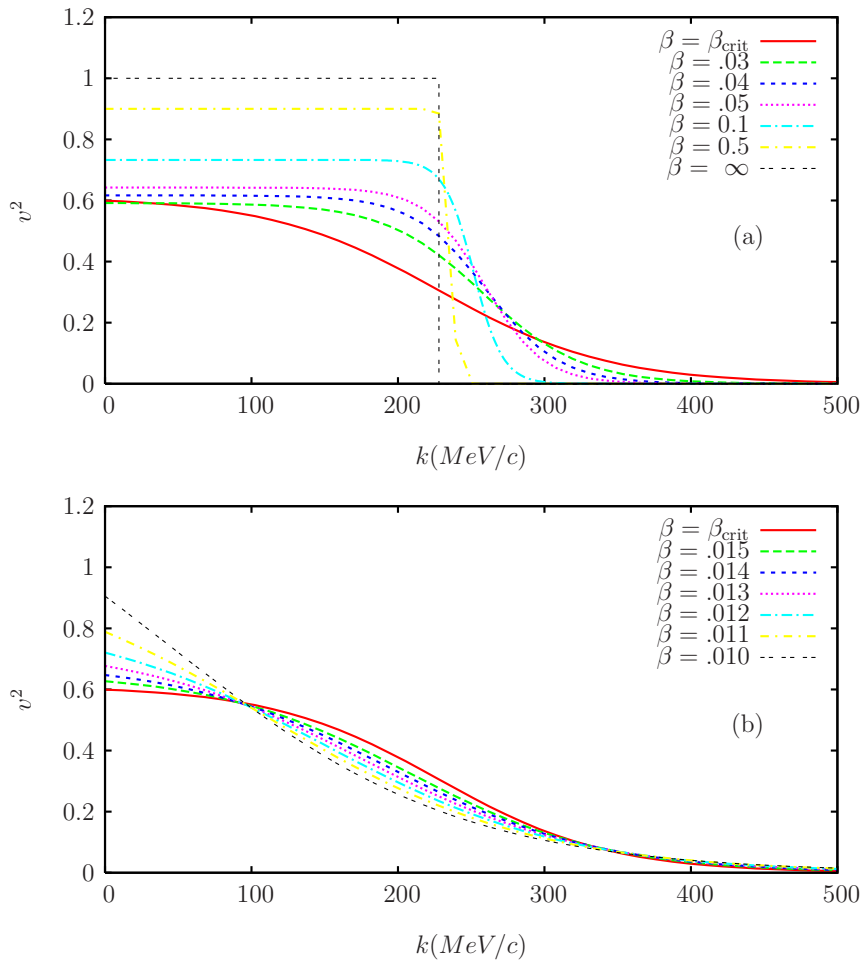}
\caption{(Color online) 
Momentum distribution of the initial state, Eq.~(\ref{eq:v2}),
evaluated for $k_F=228$ MeV/c, $\rho_A=k_F^3/(6\pi^2)$ and
different values of $\beta$ (in c/MeV) above (a) and
below (b) the critical value $\beta_{\rm crit}$.}
\end{figure}

\begin{figure}
\label{fig:fig4}
\includegraphics[scale=0.8]{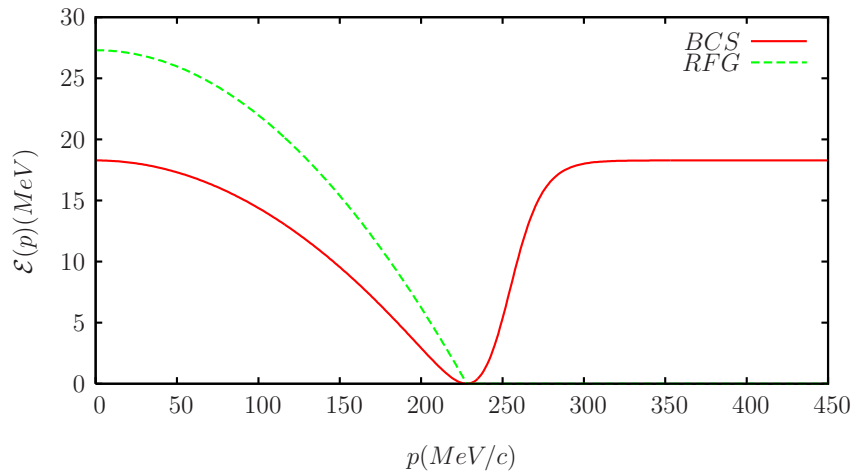}
\caption{(Color online) The excitation energy $\cal E$ computed according to
Eq.~\eqref{eq:cale} neglecting the last two terms for $k_F=228$
MeV/c, $\rho_A=k_F^3/(6\pi^2)$, $\beta=0.1$ c/MeV. The RFG results
are also shown for comparison. }
\end{figure}

The next important issue to be addressed is to determine where the
spectral function is nonzero in the $({\cal E},p)$ plane. The
answer is found in Fig.~3 where the support of the spectral
functions of the RFG and of our BCS-inspired model are displayed
and compared for $\beta$=0.1 c/MeV and the same $k_F$ used in
Fig.~1. Both spectral functions of course are just
$\delta$-functions, but concerning their support two major
differences distinguish the two:
\begin{enumerate}
\item In the range of momenta where both exist the excitation
spectrum of the daughter system is substantially softer than the
RFG one; \item For missing momenta larger than $k_F$ the BCS case,
unlike the RFG, continues to display a spectrum (its spectral
function continues to have a support) which in the thermodynamic
limit rises quite suddenly with $p$ until it reaches the value
${\cal E}$ assumes for vanishing missing momentum, namely
\begin{equation}
{\cal E}_{\rm max} = T_F |v(k_F)|^2~.
\end{equation}
This energy, which can be shown to be the upper limit of the
daughter spectrum and which cannot be exceeded because of the
stability condition in Eq.~\eqref{eq:stab}, is reached only at
$p=\infty$, but over a large span of momenta ${\cal E}$ remains
almost constant, thus corresponding to the situation of an
eigenvalue with infinite degeneracy stemming from the symmetry
$U(1)$ associated with the particle number conservation. As $p$ is
lowered, approaching the Fermi surface, the degeneracy is lifted
and we face a situation of a spontaneously broken symmetry,
reflected in the structure of our state which contains components
of all possible particle number. This situation is strongly
reminiscent of superconductivity, where the spontaneous symmetry
breaking also occurs in the proximity of the Fermi surface (in
this case, however, the control parameter is the temperature; in
our case it is the momentum itself).
\end{enumerate}

\begin{figure}
\label{fig:fig5}
\includegraphics[scale=0.8]{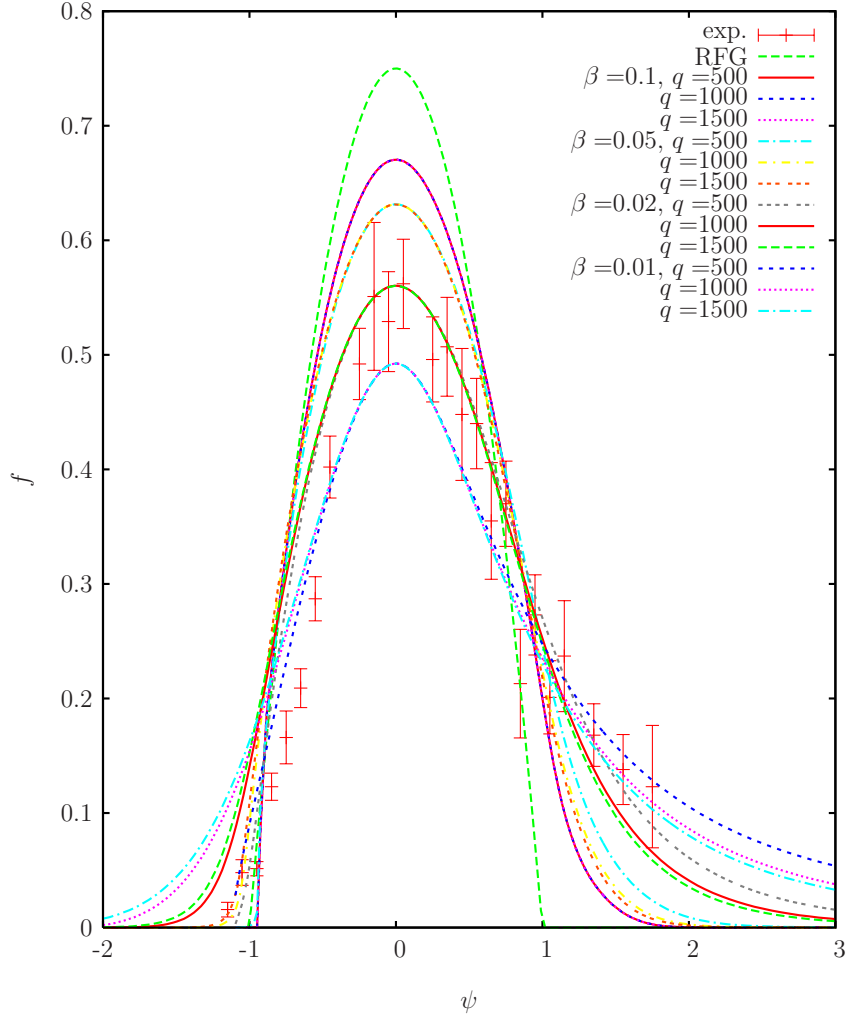}
\caption{(Color online) 
The superscaling function $f$ defined in Eq.(\ref{eq:f})
plotted versus the scaling variable (\ref{eq:psi}) in the RFG model
and in the present BCS model for three values of $q$ (in MeV/c) and
different values of $\beta$ (in c/MeV). As usual, $k_F=$228 MeV/c.
Data are taken from \cite{Maieron:2001it,JJ}.  }
\end{figure}
This set of degenerate states has a dramatic impact on the
superscaling function of our model, which we obtain using the
procedure already outlined for the RFG in Section~\ref{sec:SF}
employing the same dividing factor $\Lambda$ given in
Eq.~\eqref{eq:K}. The function $f$ is displayed versus the scaling
variable $\psi$ in Eq.~\eqref{eq:psi} for a few values of $\beta$
and $q$ in Fig.~4. For comparison the RFG result in
Eq.~\eqref{eq:fRFG} and the averaged experimental
data~\cite{Maieron:2001it,JJ} are also shown. One sees that to get
$f$ for large positive $\psi$ we have to integrate in the $({\cal
E},p)$ plane in domains (see CDM) encompassing large fractions of
those degenerate states discussed above. These are thus the cause of
the asymmetry of the scaling function with respect to $\psi=0$
appearing in Fig.~4. For $\psi$ large and negative these states are
to a large extent excluded from entering into the building up of
$f$. The fact that this effect is more and more pronounced as
$\beta$ becomes smaller reflects the impact of the tail of the
momentum distribution which indeed grows when $\beta$ decreases and,
as a consequence, more degenerate states participate to build up
$f$. Note that values of $\beta$ around the critical value
($\beta\simeq 0.01-0.02$ c/MeV) yield a tail which is in qualitative
agreement with the experimental data.

\begin{figure}
\label{fig:fig6}
\includegraphics[scale=0.8]{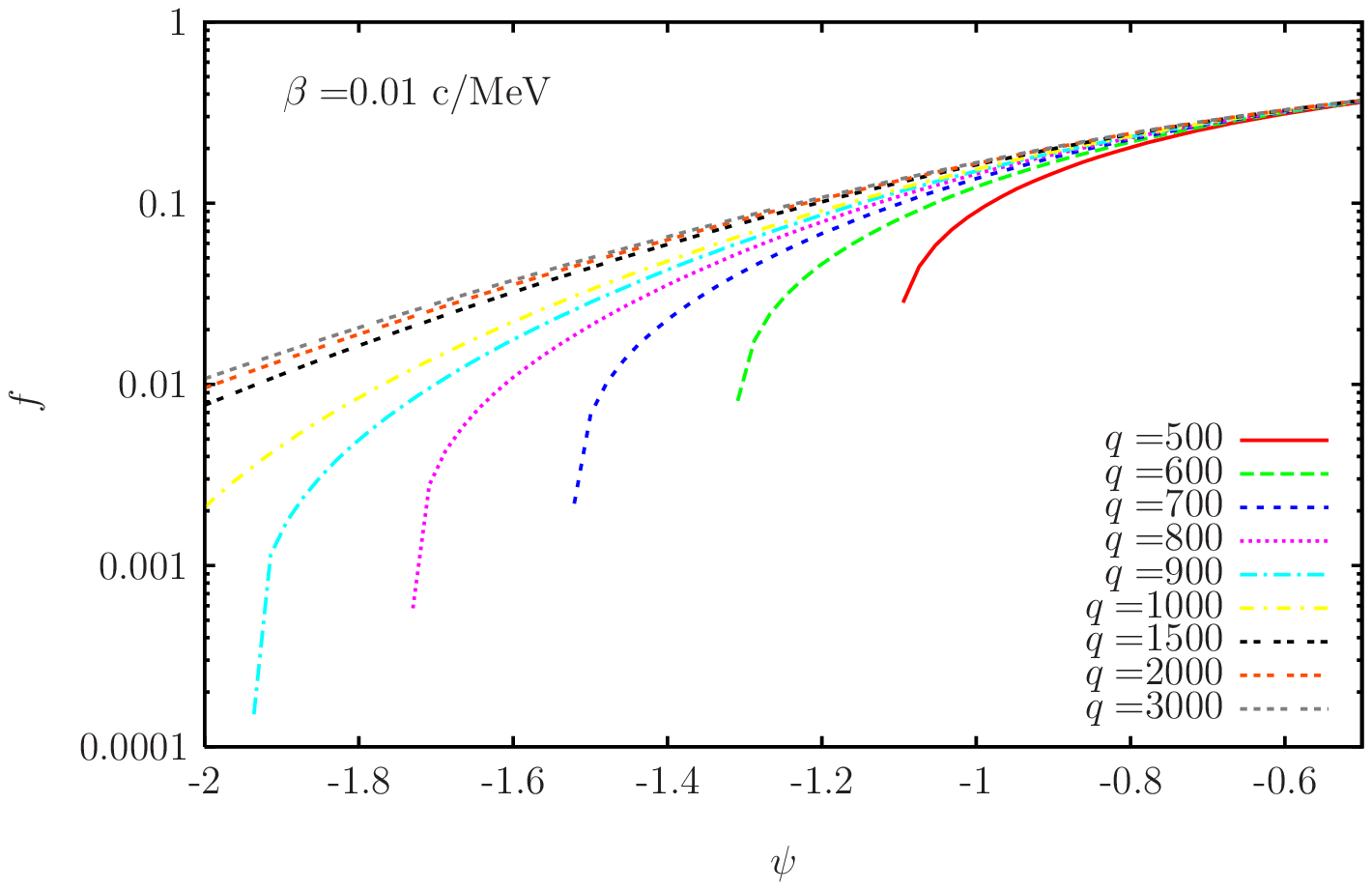}
\caption{(Color online) 
The superscaling function $f$ in the negative $\psi$
region plotted for several values of $q$ (in MeV/c) and
$\beta=0.01$ c/MeV. As usual, $k_F=$228 MeV/c.}
\end{figure}
\begin{figure}
\label{fig:fig7}
\includegraphics[scale=0.8]{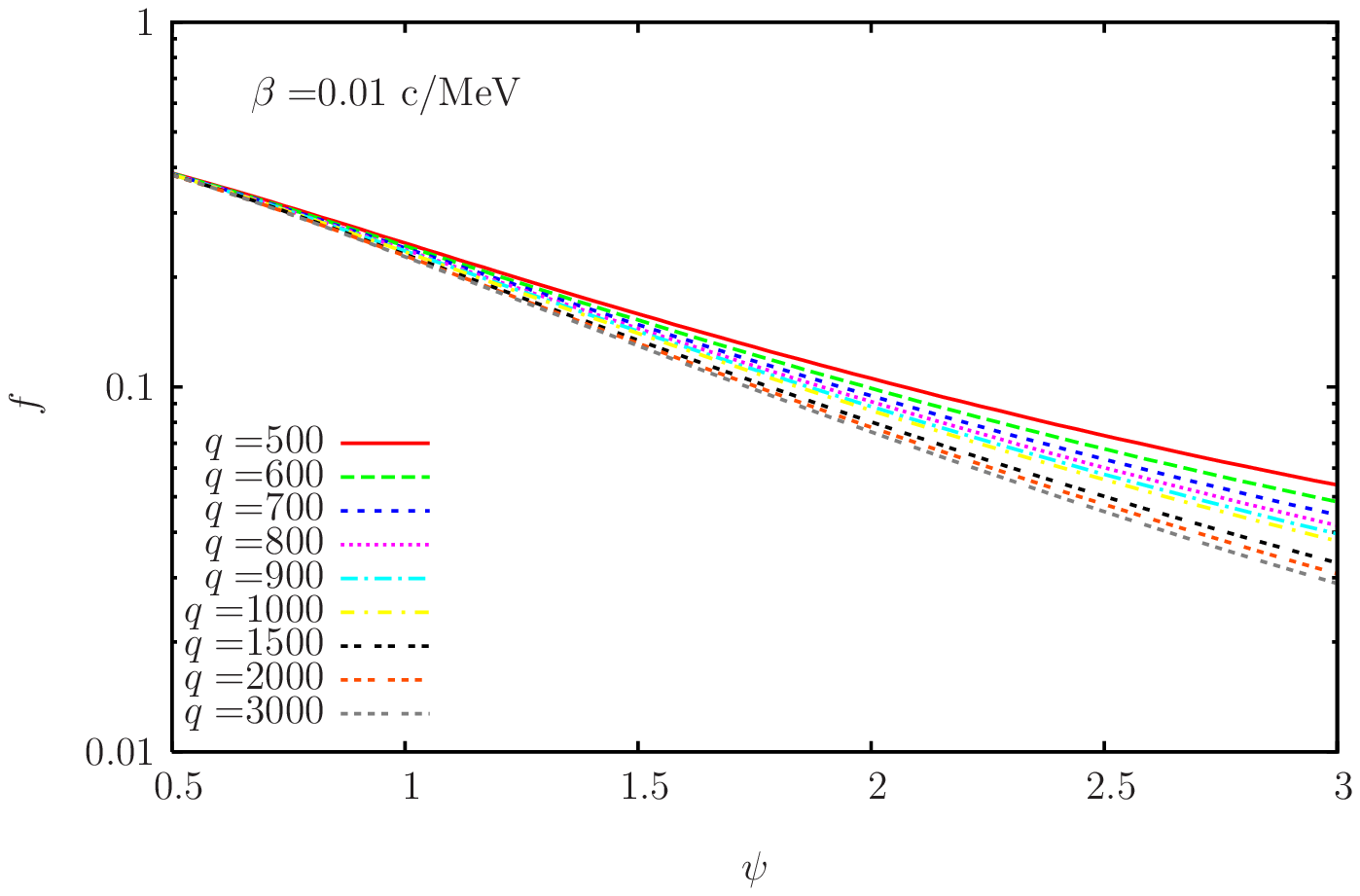}
\caption{(Color online) 
The superscaling function $f$ in the positive $\psi$
region plotted for several values of $q$ (in MeV/c) and
$\beta=0.01$ c/MeV. As usual, $k_F=$228 MeV/c.}
\end{figure}

As far as scaling of the first kind is concerned, Fig.~4 shows that
this is quickly reached in the vicinity of the QEP, although not so
to the right and to the left of it. A closer examination of the
results (see Figs.~5 and 6, where $f$ is plotted on a logarithmic
scale for a wider $q$-range at $\beta$=0.01 c/MeV) shows, however,
that also here the BCS model does scale, however with an onset
reached only for $q\simeq$ 1.5 GeV/c, namely for larger momenta than
when at the QEP where the onset already occurs at about 500 MeV/c.
Also from Figs.~5 and 6 it appears that the scaling regime is
reached faster to the right than to the left of the QEP. Moreover,
the asymptotic value for $\psi<0$ is approached from below, namely
the superscaling function grows with $q$ until it reaches its
asymptotic value, in contrast with the experimental findings. This
reflects the fact that our model, although appealingly simple, is
not able to account for features of this kind. We conjecture that a
non-uniform strength function should be associated with the
continuum spectrum of the daughter nucleus, which in general implies
the development of a more elaborate version of our model. For
instance, a further extension of our model could allow the
coefficients $v$ to become complex to account for the finite
lifetime of the excited states. Note that the same trend of
approaching first-kind scaling from below is also found in
~\cite{bulgari1,bulgari2,bulgari3} within the framework of the
Coherent Density Fluctuation Model where realistic nucleon momentum
and density distributions are
used~\cite{bulgari1,bulgari2,bulgari3}. On the other hand, in
relativistic mean-field theory~\cite{Caballero:2005sj} the approach
is from above, and thus in better accord with the experimental data.

\begin{figure}
\label{fig:fig8}
\includegraphics[scale=0.8]{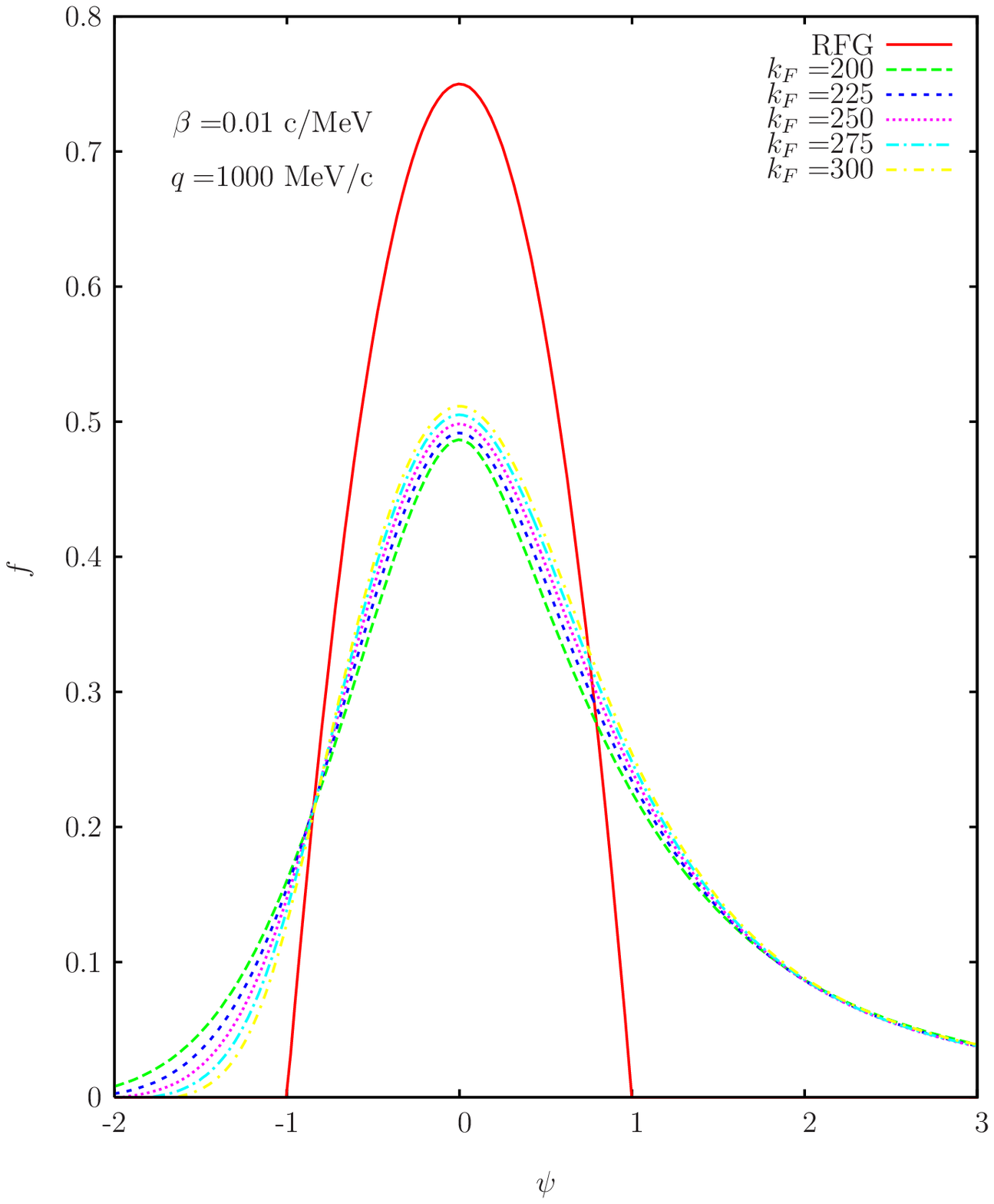}
\caption{(Color online) 
The superscaling function $f$ plotted versus $\psi$ for
several values of the Fermi momentum $k_F$ (in MeV/c) at
$\beta=$0.01 c/MeV and $q=$1000 MeV/c.}
\end{figure}

Finally, using the present BCS model, we investigate the second-kind
scaling behaviour by repeating our calculations for different values
of the external parameter $k_F$, each one taken to represent a
specific nucleus. The results are shown in Fig.~7, where it appears
that the BCS model breaks scaling of the second kind both at the QEP
and for negative $\psi$, whereas it fulfills it at high positive
values of the scaling variable. However, this is not how the
original idea of scaling of the second kind was
introduced~\cite{DS199,DS299}. In that study of experimental data a
momentum $k_A$ (actually called $k_F$ in that work, although it was
in fact a phenomenological parameter and not necessarily the
``true'' Fermi momentum as it must reflect both initial- and
final-state interaction effects) was chosen for each nuclear species
and used in the definition in Eq.~\eqref{eq:psi} of the scaling
variable $\psi$ and of the dividing factor in Eq.~\eqref{eq:K},
which will now be written as
\begin{equation}
\Lambda_A =
\frac{1}{k_A}\left(\frac{m_N}{q}\right)\left(\frac{2m_N T_F}{k_F^2}\right)~.
\end{equation}
For simplicity, in the present approach the value of $k_A$ is chosen
in order to have all the corresponding superscaling functions
coincide at the QEP, thus realizing superscaling at least where the
nuclear response is the largest. We display the results in Fig.~8,
where each curve corresponds to given $k_F$ and $k_A$, for $q=1000$
MeV/c and $\beta=0.01$ c/MeV. In this case it turns out that scaling
of second kind at the QEP can indeed be achieved using the following
empirical expression
\begin{equation}
\frac{k_A}{k_F} = 1 + \alpha (k_0-k_F)~, \label{eq:kA}
\end{equation}
where $k_0$ is just a parameter (here $k_0 \equiv k_F = 228$ MeV/c)
chosen to fix the height of the peak and where $\alpha=6\times
10^{-4}$ c/MeV. That is, $k_A$ is found to fall very slightly with
increasing $k_F$ for realistic values of the Fermi momentum.
Over much of the range of $\psi$ shown in the figure
one sees relatively good second-kind scaling, although the results
still point to a sizable violation of the second kind scaling in the
scaling domain (large negative scaling variable).
\begin{figure}
\label{fig:fig9}
\includegraphics[scale=0.8]{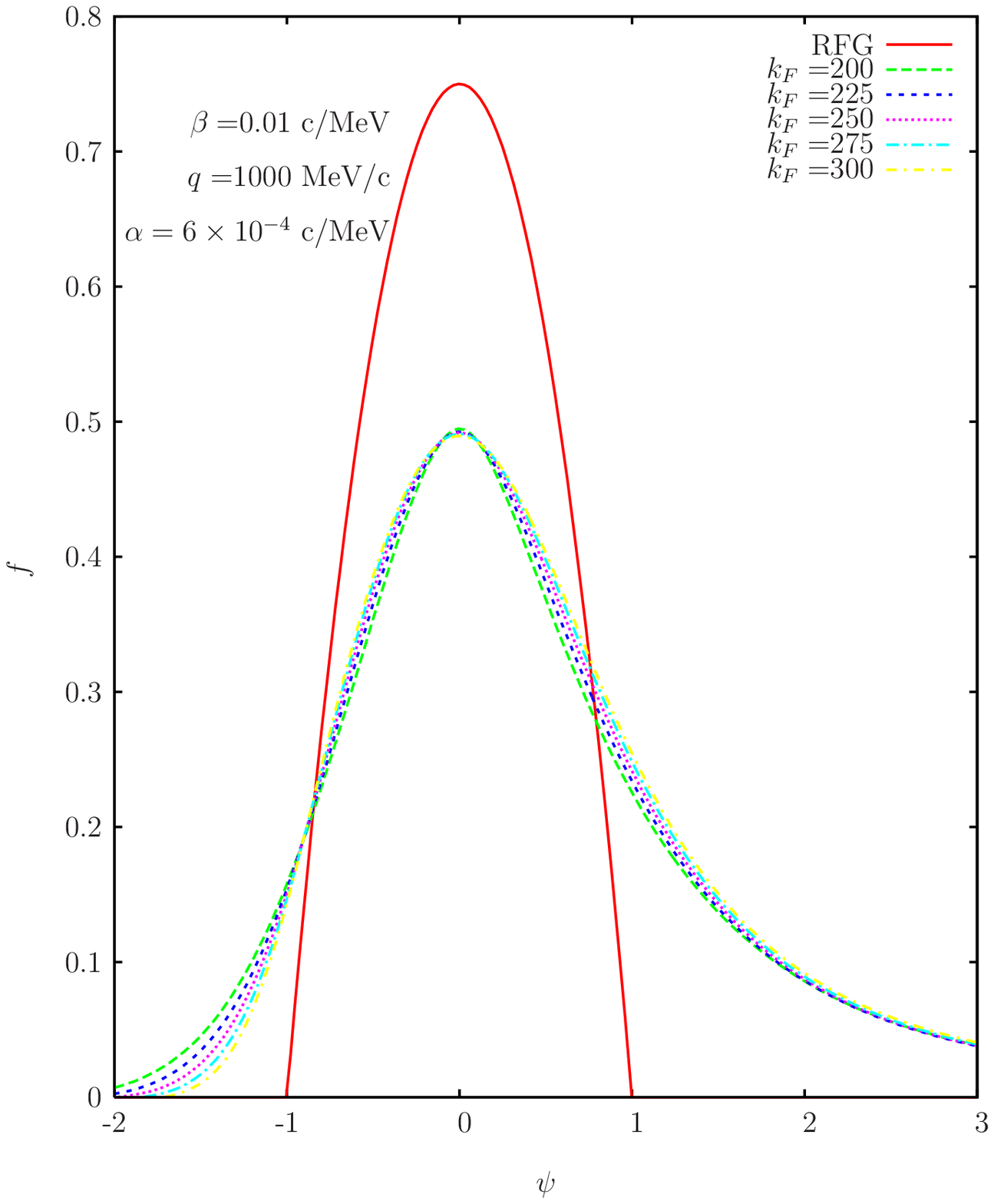}
\caption{(Color online) 
The superscaling function $f$ plotted versus $\psi$ for
several values of the Fermi momentum $k_F$ (in MeV/c) and with $k_A$
devised in such a way that the peaks coincide (see text).}
\end{figure}
\begin{figure}
\label{fig:fig10}
\includegraphics[scale=0.8]{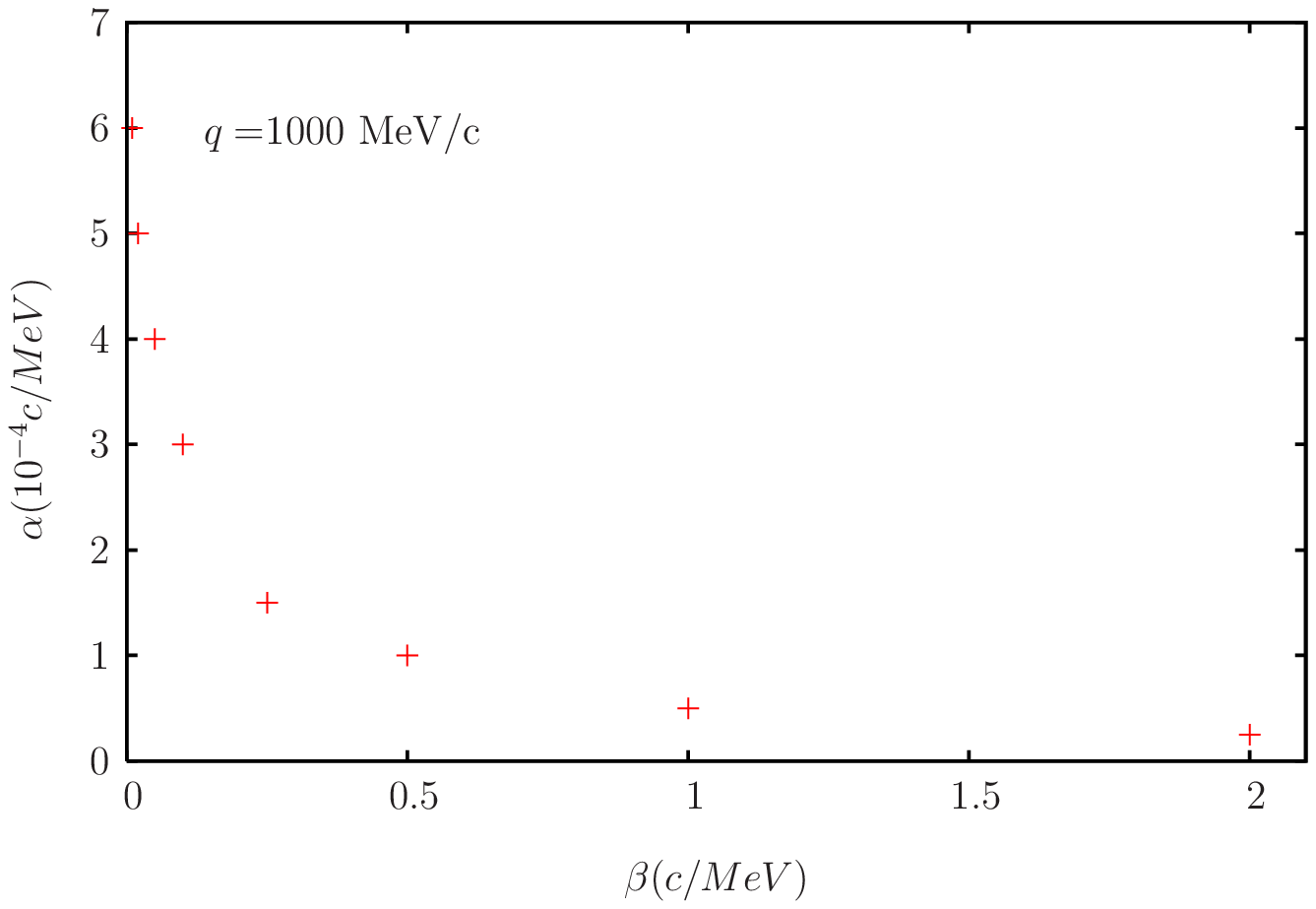}
\caption{(Color online) 
The parameter $\alpha$ of Eq.~\eqref{eq:kA} shown as a function
of $\beta$ for $q=1000$ MeV/c.}
\end{figure}
Finally, we note that for higher values of $\beta$ the parameter
$\alpha$ is found to decrease, as it should, since the RFG result
$\alpha=0$ must be recovered when $\beta \to \infty$. The behaviour
of $\alpha$ versus the parameter $\beta$ is shown in Fig.~9.


\section{Conclusions}
\label{sec:concl}

In the present study a simple extension of the relativistic Fermi
gas model for studies of relatively high-energy inclusive
electroweak cross sections has been developed. Starting from the RFG
in which a degenerate gas of nucleons is assumed for the nuclear
ground state, in this extension pairs of particles are promoted from
below the Fermi surface to above, yielding a spectral function and
the resulting momentum distribution with Fourier components for all
values of momentum. In the spirit of the RFG this new model has been
constructed in a way that maintains covariance --- the pairs of
particles involved all have spinor products with opposing momenta
and helicities, thereby producing a ground state with net momentum
zero and net spin zero.

Specifically, the model employed is inspired by the BCS approach
used in treating superconductivity, both in condensed matter physics
and in the nuclear physics problem of superconducting nuclei, in
which an infinite product of operators of the form $u_p + v_p
a^\dagger_{p\uparrow}a^\dagger_{-p\downarrow}$ with quasiparticle
weighting factors $u_p$ and $v_p$ is assumed. The specific choice of
these weighting factors characterizes the particular model assumed
in the present study, namely, such that the ground-state momentum
distribution has a rounded Fermi surface and a long tail to
represent effects from both long- and short-range correlations. It
is, however, worth recalling that our state only accounts for
correlations between pairs of fermions in time-reversed states,
namely with opposite momenta, as occur when employing the pairing
Hamiltonian. Clearly the parametrization used in the present study
is only one choice for the weighting factors and it is
straightforward to employ other more realistic forms when warranted
(see for instance \cite{Stringari90,Moya91,CiofidegliAtti:1995qe}).
The nuclear particle number and total energy of the ground state are
obtained as expectation values using this form and constrained to
chosen values $A$ and $M_A$. The final state (the one resulting
after the electroweak interaction with the ground state has taken
place) is chosen to be a plane-wave outgoing nucleon plus a daughter
nucleus recoiling in a way that conserves momentum. In the rest
frame of a general daughter state the particle number and total
energy may be found and accordingly the daughter ground state may be
identified. With these ingredients in hand it is then
straightforward to obtain the nuclear spectral function and from it
to compute the rest of the desired functions, namely, the nuclear
responses (in the present work only the EM longitudinal response is
discussed, although it is clearly straightforward to obtain any
electroweak responses using the same ideas), and the superscaling
function.

With respect to the last, the results presented in
Section~\ref{sec:results} show that the model is capable of
accounting for several, but not all, of the features seen in other
studies of superscaling, both phenomenologically and within the
context of alternative modeling.  Specifically, scaling of the first
kind (independence of $q$) is observed for momentum transfers that
are high enough. Near the quasielastic peak the onset of first-kind
scaling is quite rapid, whereas away from the peak it is somewhat
slower, that is, requires a high value of $q$ to occur. As with most
models (the exception being relativistic mean-field theory and its
derivatives) the approach to first-kind scaling is from below,
whereas the data in the scaling region approach from above. Finally,
second-kind scaling (independence of nuclear species) is seen to be
good although some scaling violations are observed.


\appendix

\section{Momentum distribution of the daughter nucleus}
\label{sec:appB}

The momentum distribution associated with the daughter nucleus
reads
\begin{multline}
  n_{D(q)}(k)=<D(q)|\sum_sa^\dagger_{ks}a_{ks}|D(q)>=\\
  \frac{1}{|\vp_q(q)|^2}\sum_s\Bigl\{<0|[\up_q(q)^*+\vp_q(q)^* a_{-q\downarrow}
a_{q\uparrow}]
  [\up_k(q)^*+\vp_k(q)^* a_{-k\downarrow} a_{k\uparrow}]
a^\dagger_{q\uparrow} a^\dagger_{ks}\times\\
  \prod_{p\not=q\not=k}[\up_p(q)^*+\vp_p(q)^* a_{-p\downarrow} a_{p\uparrow}]
  [\up_p(q)+\vp_p(q) a^\dagger_{p\uparrow} a^\dagger_{-p\downarrow}]
a_{ks} a_{q\downarrow}\\
  [\up_k(q)+\vp_k(q) a^\dagger_{k\uparrow}a^\dagger_{-k\downarrow}]
  [\up_q(q)+\vp_q(q) a^\dagger_{q\uparrow} a^\dagger_{-q\downarrow}]|0>\Bigr\}=\\
  \frac{1}{|\vp_q(q)|^2}\Bigl\{|\vp_q(q)|^2|\vp_k(q)|^2 <0|
  \prod_{p\not=q\not=k}[\up_p(q)^*+\vp_p(q)^* a_{p\uparrow}a_{-p\downarrow}]
  [\up_p(q)+\vp_p(q) a^\dagger_{p\uparrow} a^\dagger_{-p\downarrow}]|0>\Bigr\}\\
  =|\vp_k(q)|^2 (1-\delta_{kq}).
\end{multline}


\section{Evaluation of the matrix element}
\label{sec:appC}

Considering the matrix element
\begin{equation}
{\cal M}_p(q) = <D(q)|a_{p\uparrow}|BCS>
\end{equation}
we observe that
\begin{eqnarray}
a_{p\uparrow}|BCS> &=&
a_{p\uparrow}
\prod_{k}(u_k+v_k a^\dagger_{k\uparrow}a^\dagger_{-k\downarrow})|0>
\nonumber
\\
&=&\left[\prod_{k\not=p}\left(u_k+v_k a^\dagger_{k\uparrow}a^\dagger_{-k\downarrow}\right)\right]
a_{p\uparrow}\left(u_p+v_p a^\dagger_{p\uparrow}a^\dagger_{-p\downarrow}\right)|0>
\nonumber
\\
&=&\left[\prod_{k\not=p}\left(u_k+v_k a^\dagger_{k\uparrow}a^\dagger_{-k\downarrow}\right)\right]
v_p a^\dagger_{-p\downarrow}|0>
\nonumber
\\
&=& v_p a^\dagger_{-p\downarrow}
\prod_{k\not=p}\left(u_k+v_k a^\dagger_{k\uparrow}a^\dagger_{-k\downarrow}\right)|0>
\nonumber
\\
&=&v_p a^\dagger_{-p\downarrow} |BCS(\not=p)>~,
\end{eqnarray}
having defined
\begin{equation}
|BCS(\not=p)>=\prod_{k\not=p}\left(u_k+v_k
a^\dagger_{k\uparrow}a^\dagger_{-k\downarrow}\right)|0>\, ,
\end{equation}
the BCS wavefunction without the term $p$. Similarly
\begin{equation}
|D(q)>=\frac{1}{\sqrt{|\vp_q(q)|^2}}a_{q\uparrow}|BCS^\prime>
=\sqrt{\frac{\vp_q(q)}{\vp_q(q)^*}}
 \ a^\dagger_{-q\downarrow} |BCS^\prime(\not=q)>
\end{equation}
and therefore
\begin{multline}
\label{eq:B1}
{\cal M}_p(q) = v_p\sqrt{\frac{\vp_q(q)}{\vp_q(q)^*}}
<BCS^\prime(\not=q)|a_{-q\downarrow} a^\dagger_{-p\downarrow}|BCS(\not=p)>
\\
= v_p \sqrt{\frac{\vp_q(q)}{\vp_q(q)^*}}
\left(
\delta_{pq}<BCS^\prime(\not=q)|BCS(\not=p)>
-<BCS^\prime(\not=q)|a^\dagger_{-p\downarrow}a_{-q\downarrow} |BCS(\not=p)>
\right)~.
\end{multline}
Now
\begin{equation}
\label{eq:B2}
\delta_{pq}<BCS^\prime(\not=q)|BCS(\not=p)>=\delta_{pq}\prod_{k\not=q}[\up_k(q)^*
u_k+\vp_k(q)^* v_k]\, ,
\end{equation}
whereas the second term in Eq.~(\ref{eq:B1}) vanishes, since
\begin{equation}
a_{-q\downarrow} |BCS(\not=p)>=0\ \ \ \ \ \ {\rm if} \ p=q
\end{equation}
and
\begin{equation}
a_{-q\downarrow} |BCS(\not=p)>=-v_q a^\dagger_{q\uparrow} |BCS(\not=p,\not=q)>\ \ \ \ \
{\rm if}\ p\not= q,
\end{equation}
namely
\begin{equation}
a_{-q\downarrow} |BCS(\not=p)>=(\delta_{pq}-1)v_q a^\dagger_{q\uparrow} |BCS(\not=p,\not=q)>
\end{equation}
and, similarly
\begin{equation}
a_{-p\downarrow} |BCS^\prime(\not=q)>=(\delta_{pq}-1)\vp_p(q) a^\dagger_{p\uparrow} |BCS^\prime(\not=p,\not=q)>~.
\end{equation}
Then
\begin{multline}
<BCS^\prime(\not=q)|a_{-q\downarrow} a^\dagger_{-p\downarrow}|BCS(\not=p)>
\\=(\delta_{pq}-1) \vp_p(q)^* v_q
<BCS^\prime(\not=p,\not=q)| a_{p\uparrow} a^\dagger_{q\uparrow} |BCS(\not=p,\not=q)>=0~.
\label{eq:B3}
\end{multline}
Eqs.(\ref{eq:B2}) and (\ref{eq:B3}), when inserted into
Eq.~(\ref{eq:B1}), finally yield
\begin{equation}
{\cal M}_p(q)
=\delta_{pq} v_q \sqrt{\frac{\vp_q(q)}{\vp_q(q)^*}}
\prod_{k\not=q}[\up_k(q)^* u_k+\vp_k(q)^* v_k]~.
\end{equation}


\section*{Acknowledgements}
We would like to thank A.~Antonov, J.~A.~Caballero and E.~Moya de Guerra
for fruitful
discussions. This work was supported in part (TWD) by the U.S.
Department of Energy under contract No. DE-FG02-94ER40818 and in
part by the INFN-MIT ``Bruno Rossi'' Exchange program (RC and TWD).




\end{document}